\documentclass[12pt,preprint]{aastex}
\usepackage{graphicx}
\newcommand{\kms}{km~s$^{-1}$}

\newcommand{\etal}{et al.}
\begin{document}
\slugcomment{Accepted for publication: The Astrophysical Journal}
\title{Direct Evidence for an Enhancement of 
Helium  in Giant Stars in Omega Centauri\footnote{Data 
presented herein were obtained 
at the Gemini 
Observatory, which is operated by the 
Association of Universities for Research in Astronomy, Inc., 
under a cooperative agreement 
with the NSF on behalf of the Gemini partnership: the 
National Science Foundation (United States), the Science and 
Technology Facilities Council (United Kingdom), the 
National Research Council (Canada), CONICYT (Chile), 
the Australian Research Council (Australia), 
Minist\'{e}rio da Ci\^{e}ncia e Tecnologia (Brazil) 
and Ministerio de Ciencia, 
Tecnolog\'{i}a e Innovaci\'{o}n Productiva (Argentina).
This paper also includes spectra gathered with the 6.5-meter Magellan
Telescope/CLAY 
located at Las Campanas Observatory, Chile.}}

\author{A. K. Dupree}  
\affil{Harvard-Smithsonian Center for Astrophysics, 60 Garden Street,
  Cambridge, MA 02138}
 \email{dupree@cfa.harvard.edu}

\author{Jay Strader\footnote{Hubble Fellow, now Menzel Fellow at
    Harvard College Observatory, Cambridge, MA USA.}}
\affil{Harvard-Smithsonian Center for Astrophysics, 60 Garden Street,
  Cambridge, MA 02138}
\email{jstrader@cfa.harvard.edu}

\author{Graeme H. Smith}
\affil{University of California Observatories/Lick Observatory, 
 University of California, Santa Cruz, CA 95064}
\email{graeme@ucolick.org}

\begin{abstract}
The double main sequence identified in the globular cluster Omega
Centauri has been interpreted using isochrones to 
indicate a large variation in the abundance of helium.
If true, a helium enhancement carries strong implications for the
chemical and stellar evolutionary history of this cluster. 
However, only indirect measures currently  support  this conjecture.
We report the discovery of a variation
in the line strength of the near-infrared \ion{He}{1}
10830\AA\ transition in twelve similar red giants in Omega Centauri observed with
PHOENIX on Gemini-S.  Abundances of these stars 
derived from Magellan/MIKE spectra taken at Las Campanas Observatory
show that the  helium transition is not
detected in the most metal-poor population ([Fe/H] $<$ $-$1.8), yet is present in the majority of stars
with [Fe/H] $\geq$ $-$1.8. These observations give the first direct evidence for 
an enhancement of  helium  in Omega Centauri.
The appearance of helium appears  better correlated with  increased [Al/Fe]
and [Na/Fe] abundances  than as a function of [Fe/H], giving
observational support to the presence of high-temperature H-burning in
a prior generation of stars.

\end{abstract}


\section{Introduction}

Omega Centauri ($\omega$ Cen) is a globular cluster long known
to have a color spread along the red giant branch (Woolley 1966; Geyer
1967) indicative of a spread in the [Fe/H] abundances as originally
identified in cluster RR Lyrae variables (Freeman \&\ Rodgers 1975).
Among subgiants too, four (and possibly five) subgroups of different ages and
metallicities have been identified (Villanova \etal\ 2007; Bellini \etal\ 2010).
The discovery (Anderson 1997; Bedin et al. 2004)  of multiple stellar 
populations on the main sequence in a globular cluster, first
identified in $\omega$ Cen was possibly 
a singular anomaly. But another globular cluster, NGC 2808, displays multiple branches
on  the  main sequence  (Piotto \etal\ 2007).   These results, based on unevolved dwarf stars,
present firm evidence for multiple episodes of star formation in a
globular cluster, and  upset the long-held 
paradigm that a globular cluster consists
of a coeval population of stars (Piotto 2009). 

Norris (2004) offered an explanation for the 'double' main sequence in
Omega Cen based on stellar models, as resulting from a 
substantially enhanced helium abundance in the 'blue' main sequence 
($\Delta$Y=+0.13; Y=0.38), over the primordial abundance
(Y=0.25) in the 'red' sequence.  Subsequently, the blue main-sequence 
of stars in $\omega$ Cen was discovered to be more metal-rich 
than the red main sequence ([M/H]$_{blue}$=$-$1.26 vs
[M/H]$_{red}$=$-$1.57; Piotto \etal\ 2005). The cluster NGC 2808 shows a similar 
effect (Bragaglia \etal\ 2010a).

An enhanced helium abundance would have a number of ramifications.
Stellar models suggest that  excess helium could account for many characteristics of
globular cluster color-magnitude diagrams:  
an extended horizontal branch (HB), gaps in the HB, 
RR Lyr periods and period distributions, and the ratio 
of blue to red HB stars (D'Antona \& Caloi 2004, 2008; Norris 2004;
Catelan 2009; Gratton \etal\ 2010).
It is currently thought that a helium
enhancement and other chemical inhomogeneities might result
from self-enrichment (pollution) of a second generation of cluster stars
by the processed material from the first generation stars. Various sources
for excess helium have been proposed: production by first generation asymptotic branch
stars (AGB) of intermediate mass,  or massive hot stars; and Type II supernovae.  
However models 
can not uniquely  identify  the source of the  enrichment;
concerns remain because {\it all} of the processed first generation
material is required to produce enrichment, and the material from large numbers of stars
must be homogenized (Norris 2004; Piotto et al. 2005).  Omega Cen
appears to have an anomalously blue HB (D'Cruz \etal\ 2000; Momany 
\etal\ 2004) which has been attributed (Norris 2004; Piotto \etal\ 2005)
to the progeny of the blue main sequence.
However,  AGB models (Choi \& Yi 2008) can not produce the 
quantity of helium in first generation stars as required for Omega
Cen. 

While there is great interest in the possibility of
a helium enhancement in these multiple populations, only 
indirect methods have been used to date to infer this enhancement 
in globular cluster stars. The observable quantities rely principally on the appearance
of the color magnitude diagrams,  comparison with
stellar isochrones,  and the composition of elements 
other than helium, in particular Na and Al, which signal the products of high-temperature 
H-burning (Densikov \& Denisenkova 1989; Langer \etal\ 1993) including
CNO, NeNa, and MgAl cycles that lead to helium production (cf. Gratton \etal\ 2001; Gratton \etal\
2004; Bragaglia et al. 2010b).

Helium is a difficult element to detect spectroscopically 
in cool stars because high temperatures are generally required to 
produce the optical transitions.  Hot stars, such as the HB objects,
where helium might appear, are subject to diffusion and preferential
settling of elements making the determination of abundances ambiguous 
(Da Costa \etal\ 1986; Behr \etal\ 1999).  However, the infrared \ion{He}{1} 
transition at 10830\AA, has been detected in metal-poor red 
giant stars in the field (Dupree et al. 1992, 2009) as well
as in the globular cluster M13 (Smith et al. 2004).
The strength of the helium absorption is independent of [Fe/H] metallicity
in the field-star sample (Dupree \etal\ 2009).  A calculation of the 10830\AA\
profile in the quiet sun with PANDORA (Avrett \& Loeser 2008) demonstrates
that a change in the helium abundance from Y=0.28 to Y=0.4 increases
the equivalent width of the absorption by a factor of 3. Thus a simple measure of
the distribution of equivalent widths in stars within a restricted region  of
temperature and luminosity could reveal a variation
in helium strength. We report the results of such a study here of
stars in the globular cluster $\omega$~Cen.  Because this cluster is believed to have
substantial helium enhancement (Norris 2004), it is an optimum target.

Spectroscopic observations of $\omega$ Cen giants at Gemini-S (for the
near-IR \ion{He}{1} transition) 
and Magellan/Clay (for optical spectra) are
presented in Section 2; the derivation of abundances from Magellan spectra is
addressed in Section 3; results are discussed in Section 4, and final
comments in Section 5.

\section{Observations and Reduction}
Selection of target stars in $\omega$ Cen 
was guided by the fact that the \ion{He}{1} $\lambda$10830 transition appears
in metal-poor field giants with $T_{eff} \ge 4500K$ where the stellar
chromospheres reach temperatures (T $\sim$ 10,000K) 
high enough to produce the helium line (Dupree \etal\
2009).  A restricted region of the red giant branch   was
selected to fit the temperature criterion.   The target region 
spanned 0.05 in $B-V$ color and  was limited to a range 
of $\sim$0.5 magnitudes in V between 12.95
and 13.48 (Fig. 1).  This selection avoids asymptotic giant branch stars 
and minimizes or eliminates any temperature or luminosity dependencies of 
the line strength. In addition, horizontal branch stars
are avoided to eliminate objects where elemental diffusion might
occur.   All of the target stars have a membership probability of 
98 to 100\% based on proper motion studies (van Leeuwen \etal\ 2000),
and our radial velocity measures, discussed later,   
are consistent with membership as well.  

\subsection{Gemini-S/PHOENIX Infrared Spectra}

Classical time was awarded (Program GS-2010A-C-7) for 3 nights  on 
PHOENIX\footnote{See http://www.noao.edu/ngsc/phoenix/phoenix.html}
mounted on  Gemini-S (Hinkle \etal\ 2003)  
to obtain high-resolution spectra of the \ion{He}{1} 10830\AA\ line. 
Three clear nights of observations  were obtained 
beginning 24 March 2010 UT. 
The PHOENIX setup consisted of a slit width of 4 pixels yielding a
spectral resolution of $\sim$50,000. The order-sorting filter J9232 was
selected which spans 1.077--1.089 $\mu$m, and allows access to the
\ion{He}{1} 1.083  $\mu$m line which is shifted by $\sim$+230 \kms\ 
(equivalent to 8.3\AA) due to the radial velocity of the $\omega$~Cen cluster. 
Standard  procedures were followed by acquiring target spectra using a nodding
mode  (A-B) with a spatial separation of 5 arcsec. 
Standard stars and telluric comparison objects were observed with a nodding pattern using a
separation of 8 arcsec to avoid contamination of the targets by a residual
image. Exposure times for most of  the $\omega$~Cen giants
totaled 2 hours, and 1 hour for two of the brightest targets.    
To verify the grating setup, a spectrum of a
nearby star with strong P Cygni-type emission at 10830\AA\ was
obtained.      Because standard comparison lamps have a
sparse wavelength pattern in this near-infrared region, we observed a bright 
Population I K giant containing many securely identified 
narrow absorption lines in order to determine the wavelength scale. 
Throughout the night,  spectra
of fast-rotating B stars were obtained at various air masses  to
monitor the strength of the telluric water vapor features.  Flat
exposures and dark exposures were taken at the end of each night.  The
twelve red giants comprising the targets in $\omega$ Cen are listed in Table 1.

A technical issue with PHOENIX became apparent during our run at
Gemini-S. During the periods of good seeing, ($\sim$0.4 arcsec), the
spectrum appeared double in the spatial direction, 
leading to slightly degraded image quality.   The source of this doubling is
not now known.\footnote{The doubling phenomenon has been observed at Gemini-S since
PHOENIX began operation on Gemini-S in 2002 
(B. Rodgers, private communication). Two possibilities have been 
suggested to us: the telescope was slightly out of  focus on the slit 
(K. Hinkle, private communication) or an astigmatic collimator whose effects become 
substantially more pronounced at longer wavelengths as shown in
PHOENIX spectra obtained at the KPNO 2.1 coud\'e feed in 1998 (C. Kulesa, private
communication).  We assume this was not noticed previously since, in
queue mode, PHOENIX was not likely to be used during optimum seeing
conditions.}
Other than a moderate reduction in the S/N, this issue does not affect
any of our results.

The raw images were corrected for cosmic rays using L.A.Cosmic (van
Dokkum 2001). After this, the nodded image pairs were subtracted from each
other. These subtracted images were divided by a dark-corrected,
normalized flat field, followed by optimal extraction of one-dimensional
spectra. Owing to the paucity of arc lines in the region of the 10830
\AA\ line, the spectra were wavelength calibrated using observations of
strong-lined K giants with known radial velocities. The individual
calibrated spectra were combined with proper variance weighting. 
The spectra were normalized by excluding the region between
10833--10844\AA\ 
where the Doppler-shifted \ion{Si}{1} line ($\lambda$10827.09) 
and \ion{He}{1} line ($\lambda$10830.34) occur, and  fitting a second-order
cubic spline to the spectra.  Spectra of the helium line region in the
target giants 
are shown in Fig. 2.  

Equivalent widths of the \ion{He}{1}
absorption were measured from the unbinned data.  Spectra were not 
corrected for telluric absorption, but the equivalent width of a  
water vapor line, if blended with helium,  
was subtracted from the total helium absorption. 
Our spectra of rapidly rotating B stars, taken as standard stars, 
display a correlation between the equivalent width of the strongest 
water vapor line in this region, $\lambda_{air}$=10832.109\AA\ 
(Breckinridge \& Hall 1973) and the 10838.034\AA\ line which in some
stars blends with  the helium transition.  A straight line represents
the correlation with a $\pm$2m\AA(1$\sigma$) error of the fit. This correlation
was used to estimate the strength of 10838.034\AA\ in the target stars based on the
observed strength of 10832.109\AA.  Water vapor lines
are marked in Fig. 2 for the $\omega$ Cen giants,  and a sample of the
telluric standard stars is shown in Fig. 3. The water vapor
transition at 10838.034\AA\  varied in strength from 4 to 18 m\AA.
Where helium is not apparent in our targets,
the equivalent width in Table 1 represents an upper limit to the strength of a
non-observed helium line, and generally corresponds to the  
equivalent width of the water vapor transition. 

Radial velocities of the targets were determined from the high resolution infrared
spectra, based on fitting  the \ion{Si}{1} photospheric absorption
line  with a gaussian profile.  We also measured central wavelengths
of 10-13 \ion{Fe}{2} lines for each star from the MIKE spectra  for the abundance
analysis discussed later, and velocity measures from the individual lines have
a dispersion of less 
than 1.7 \kms. The \ion{Fe}{2}  velocities agree with
the results from \ion{Si}{1} to better than 0.2 \kms. 
Values of the radial velocity are listed in Table 1.
The mean radial velocity of $\omega$ Cen has
a distribution peaked around 
$+$233 \kms\ (Reijns \etal\ 2006). Our targets are clearly
radial velocity members of the cluster confirming the membership 
determination from proper motions (van Leeuwen \etal\ 2000). For the nine  targets in common with
Reijns \etal\ (2006), our values  agree to better than 2 \kms, with
the exception of LEID 54064 where we differ by 8.1 \kms; this star has the
largest error of 4.7 \kms\  reported by Reijns \etal\ (2006).

\subsection{Magellan/MIKE spectra}

Echelle spectra of the  PHOENIX targets were obtained at the Magellan/Clay telescope on 30 June - 2
July 2010 (UT) using the MIKE spectrograph (Bernstein \etal\ 2002) 
with a 0.70$\times$5 arcsec slit.  The spectral resolution with this
setup ranges from $\sim$26,000 on the blue side ($\lambda\lambda$3350--5000)  to $\sim$36,000 on
the red side ($\lambda\lambda$4900--9300).  Total exposure
times on individual  targets varied from 40 minutes to 1 hour, generally divided between
two exposures.  An IDL pipeline developed by S. Burles, R.
Bernstein, and J. S. Prochaska\footnote{ See
  http://www.lco.cl/telescopes-information/magellan/instruments/mike/idl-tools}
was used to extract the spectra.  This efficient pipeline evaluates
the detector gain, processes
the milky and order-tracing flats, obtains the Th-Ar arc solution, 
constructs a slit flat, and extracts a spectrum including sky subtraction.
IRAF procedures 
were used to flatten individual echelle orders, and combine multiple exposures.
The signal-to-noise ratio ranged from 100 to 150 in the combined spectra, depending on
position in the echelle blaze. Spectra were  normalized
with  a cubic spline that well fit the  standard white 
dwarf LTT 9491. After normalization,
individual spectra were combined.  Lines of \ion{Fe}{2}, \ion{Na}{1},
and \ion{Al}{1}  selected for abundance
determinations occur  in
8 echelle orders drawn from both blue and red sides of MIKE.
The equivalent widths were measured using the IRAF routine {\it splot} 
by fitting the profiles with a gaussian curve on the
continuum-normalized spectra. In some cases, where blending was present, the
blended profile was deconvolved using multiple gaussian profiles in
order to extract the equivalent width of the line.

\section{Determination of Stellar Abundances}

Abundances of Fe, Al, and Na  were derived from Magellan/MIKE spectra 
of the target stars. 
Following the precepts of Kraft and Ivans (2003), we focused on
deriving Fe abundances from \ion{Fe}{2} lines since this ion dominates
in cool giants.  The selection of \ion{Fe}{2} lines
and oscillator strengths was taken from Fulbright (2000).  Of the
26 \ion{Fe}{2} lines listed by Fulbright, we chose 10-13 to be measured
in each star, aiming to find those near the center of the echelle blaze from each 
order where the signal-to-noise is highest.

The abundance analysis was restricted to lines with reduced
equivalent width $log$(EW/$\lambda$)$<-4.5$.  In addition, we only used lines with EW $> 9$ m\AA,
unless few stronger or no lines  were available for a given species and star.
Atomic parameters for all lines measured and individual equivalent widths are
listed
in Table 2.

We derived abundances by force-fitting equivalent widths in the 2010
version of MOOG (Sneden 1973\footnote{See
  http://www.as.utexas.edu/\~{}chris/moog.html}). 
Effective temperatures for 11
of the stars were estimated using their $V-K$ colors  and the calibration of Alonso
\etal~(1999). The K-magnitudes were taken from the 2MASS All Sky
Catalog 
(Cutri \etal\ 2003) which lie within 0.005 mag of the TCS color system
adopted by Alonso \etal~(1999). One star did not have a $K$ magnitude; we used its $B-V$
color instead. A reddening of $E(B-V) = 0.12$ mag was assumed (Harris 1996). These
photometric estimates of effective temperature have small random errors:
0.01 mag in $V-K$ corresponds to a change of only 10K; 0.01 mag in
$E(B-V)$ corresponds similarly to  30K. Systematic effects dominate,
and are estimated at less than 100 K.

Surface gravities were estimated using bolometric corrections from Alonso
\etal~(1999) for the effective temperatures already derived, a distance of
$4.8\pm0.3$ kpc (van de Ven \etal~2006), and an assumed mass of 0.8
$M_{\odot}$. The gravity is insensitive to reasonable uncertainties in
mass and temperature as they enter the calculation only logarithmically.
The largest source of error is the distance, which introduces a $1\sigma$
random error of approximately 0.05 dex.

To obtain chemical abundances, Castelli \& Kurucz (2004) stellar
atmosphere models were employed. These are one-dimensional, plane-parallel models
without convective overshooting. We used $\alpha$-enhanced models
([$\alpha$/Fe] = +0.4) that are generally appropriate for stars in a
globular cluster. Models for the relevant parameters of each star were
derived through interpolation. The effective temperature and surface
gravity were fixed by the photometry as discussed above. We derived the
metallicity [Fe/H] and microturbulence $v_t$ iteratively. To minimize
non-LTE effects, only \ion{Fe}{2} lines were used. The 
microturbulence was determined in the usual manner
by requiring that there be no trend of abundance with reduced equivalent
width. The final adopted parameters for each star are given in Table
1. 

Abundances for sodium and aluminum were derived using 2--4 weak lines per
species and the stellar parameters estimated from the photometry and the
Fe II lines.  Four lines of \ion{Al}{1} were  selected: $\lambda$6696.03, $\lambda$6698.67,
$\lambda$7835.32, and $\lambda$7836.13.  The \ion{Na}{1} analysis
typically used 5 lines:
$\lambda$4668.57 (in some stars), $\lambda$5682.65, $\lambda$5688.21,
$\lambda$6154.23, 
and $\lambda$6160.75. For one star, LEID 63027, we also included the
Na  resonance lines at $\lambda$5889.96 and $\lambda$5895.94 and they 
gave results consistent with those of weaker lines. 
Wavelengths of the transitions and their {\it gf}-values were taken
from the compilation of Fulbright (2000).  The {\it gf}-values for Al
came from Jacobson \etal\ (2009).  For a subset of the stars, 
we could only derive upper limits
to the abundance of aluminum. In these cases, the reported numbers are a
median of the limits for the individual lines.
Abundances are reported relative to solar values. The solar values in
Asplund \etal~(2005) were used for all elements except for iron, for which we assume
log $\epsilon$(Fe) = 7.52 (Sneden \etal~1991).  Fig. 4 shows
a section from the  MIKE spectra of two stars similar in T$_{eff}$, {\it log g},
and [Fe/H] yet differing substantially in Na and helium.

After our analysis was completed, Johnson and Pilachowski (2010) published
abundances for many red giants in Omega Cen based on HYDRA spectra
of lower resolution [R($\lambda$/$\Delta \lambda$)=18,000].
Their sample includes nine of our targets. For the  stars in common,
our values for $T_{eff}$ differ by a mean of 9K, and a maximum of
29K. The average difference in the values of {\it log g} amounts to 0.13.
These authors derived an iron abundance generally based on
a majority of \ion{Fe}{1} lines.  Some stars had no \ion{Fe}{2};
others had one, or at most two lines from \ion{Fe}{2}  
available.  Fig. 5 shows the derived abundances for the
stars in common where an mean offset of $-$0.12$\pm$0.12  dex exists between our values
and those from Johnson \& Pilachowski. The maximum deviation of 0.2
dex from this offset is less than the scatter of 0.65 dex among other values 
commonly found in the literature (see Fig. 5 of Johnson \& Pilachowski 2010).  

\section{Results}

Helium was detected in 5 of the 12 red giants observed (see Fig. 2 and
Table 1), and the
equivalent widths of the lines differ from star to star. Considering the small range
of effective temperature and luminosity in these giants, the
equivalent width variation suggests a similar
variation in the helium abundance.  These helium lines, where
detected, have larger equivalent widths than the helium lines
found (Dupree \etal\ 2009) in metal-poor red giants with
comparable temperatures in the field (Fig. 6).  Additionally, where helium is
not found, the lines in $\omega$ Cen tend to be weaker than in the
field stars.  For the $\omega$ Cen giants, 
no correlation is apparent between position in the color-magnitude diagram (CMD)
and the presence of helium or with the iron abundance (Fig. 7).  Our sample
is small, and more targets are necessary to develop a clear pattern
between helium line strength and the iron abundance, if, indeed it 
exists.  In the metal-poor field giants, we also found no correlation
of helium strength with [Fe/H]. On the subgiant branch of $\omega$ Cen
also, Villanova \etal~(2007) found poor correlation  between [Fe/H] 
and CMD location.  The helium line, when present, 
may become stronger in the less luminous objects (Fig. 8).

\subsection{Overall Characteristics of the Presence of Helium}

The current view holds that increased helium abundance causes  the blue main sequence 
of $\omega$ Cen (Norris 2004), and the gross characteristics of the
$\omega$~Cen 
giants in the PHOENIX sample exhibiting
helium are consistent with both the fraction  and spatial distribution of
the blue  main sequence.      The
fraction of stars displaying helium, 42\%,  is comparable to the 
fraction of blue main sequence
stars (25 to 35\%) considered to have enhanced helium and also enhanced
metals (Bedin \etal\  2004; Piotto \etal\ 2005). The spatial
distribution is also of interest.  The blue main
sequence stars have a higher spatial concentration near the cluster core
than the red main sequence (Bellini et al. 2009; Sollima \etal\ 2007). Within the half-mass 
radius of the cluster ($\le$ 4.2 arcmin), the ratio of blue to red
main sequence stars ranges from 1.1 to 0.8; at 6 arcminutes
from the center the ratio is still high at 0.5; further out from the
center ($\ge$ 8 arcmin) the ratio falls to 0.4. Taking the
center of the $\omega$ Cen cluster (Anderson \& van der Marel 2010)   
as RA = 13:26:47.24, Dec=$-$47:28:46.45 (J2000.0), our sample
of twelve giants is evenly split in radius at a distance of  6 arcminutes from the center.
Of the 6 stars within 6 arcminutes from the center, 3 have
helium or 50\%; this fraction falls to 33\% (2 out of 6 targets)
for the more distant objects. While our sample is small, these
fractions are consistent 
with the spatial behavior of the blue:red ratio.
Thus, the percentage
detection and the spatial distribution 
of the helium detections are consistent with the blue main sequence population.

The equivalent widths of the detected helium lines are stronger
than the upper limits of the  non-detections by factors of 3 to 10.
Judging from our calculations for the quiet Sun, the amount of this 
enhancement could correspond to a change in helium abundance by more than a 
factor of two. However, conditions in a red giant chromosphere are certainly
different from a dwarf star.  Outflows can increase the absorption
from the metastable lines in helium (Dupree \etal\ 2009).   
Variability  in the   chromospheric structure  could
occur also, although it is not likely  to be large judging by the
lack of H-$\alpha$ wing emission,  and the weak \ion{Ca}{2}
emission reversals in the optical spectra of these giants.  Detailed
non-LTE semi-empirical calculations are underway to extract the 
helium abundance, and the chromospheric models must be constrained by
H-$\alpha$ and \ion{Ca}{2} line profiles.

\subsection{Abundances of the Giant Stars}

The relation 
between the abundance of iron [Fe/H] and the presence of helium
is shown in Fig. 9.  Stars with [Fe/H] $<$ $-$1.8 do not exhibit
helium; the majority of stars in the range 
$-1.8 \leq [Fe/H] \leq -1.4$ display the  helium transition, and the
one metal-rich star in our sample at [Fe/H]=$-$1.15 does not show
helium.\footnote{Our abundance for LEID 54084, [Fe/H]=$-$1.79, 
differs from the  Johnson \& Pilchowski (2010) value of
[Fe/H]=$-$1.5. Such a shift would not significantly change the distribution of helium
detections.}  On the main sequence, the metallicity, [M/H] corresponds to
$-$1.26 for the blue main sequence and $-$1.57 for the red main
sequence (Piotto \etal\ 2005).  An empirical transformation from [M/H]
to [Fe/H] was estimated by Villanova \etal\ (2007) from a comparison of
subgiant spectra to spectral syntheses for various values of [M/H].
The best-fitting synthesis for [M/H] was compared to values of [Fe/H] 
for many subgiants in $\omega$ Cen. The correction, obtained from the mean difference
between [M/H] and [Fe/H] amounts to $-$0.11 dex applied to values
of [M/H]. For the blue and red main sequences, this leads to  
[Fe/H]=$-$1.37 and $-$1.68 respectively. An additional 
minor red branch  can be found
on the main sequence (Villanova et al 2007; Anderson and van der Marel
2010)  termed MS-a which has higher [Fe/H] ($-$1.1 to $-$0.6) than the
blue main sequence.  We do not find  a correlation between the
[Fe/H] distribution and the presence of the \ion{He}{1} $\lambda$10830 line.

A different indication of processed material can be found in the light element
abundances of Na and Al which signal high temperature H burning via
the CNO cycle (Denisenkov \& Denisenkova 1989; Langer \etal\ 1993) in a previous 
stellar generation (Gratton \etal\ 2001).  In red giants,
the outer convective envelope will never reach the the H-burning shell
where the light elements are produced during evolution on
the red giant branch (Gratton
\etal\ 2004).  For enhanced light elements to be present, they 
must have originated in a previous generation of stars.  
The abundances of   [Na/Fe]  and [Al/Fe] are 
correlated with each other, and clearly enhanced in the stars
displaying \ion{He}{1} lines (Fig. 10). Stars without a helium
detection have low values of [Na/Fe] and principally upper limits
on the [Al/Fe] abundance.   These enhancements of Na and Al  agree with 
values from red giants in $\omega$ Cen (Norris \& Da Costa 1995;
Johnson \&
Pilachowski 2010). And the values of [Na/Fe] and [Al/Fe]  for the helium and non-helium giants  
are similar to the
abundances found in NGC 2808 for two stars, one each from the blue and 
red components of the main sequence respectively (Bragaglia \etal\ 2010a).
Thus the red giants in $\omega$ Cen exhibiting  
the abundance pattern expected from
the high temperature H-burning processes are the majority of  red giants
displaying enhanced He lines.

\subsection{Evolutionary Connections}

The population structure in $\omega$ Cen is quite complex. 
We now focus on the spectroscopic [Fe/H] values, and consider
how the helium detections fit in  with the continuity of the
color-magnitude diagram (CMD) and the various 
population components.  We follow the ideas of Villanova \etal\ (2007),
to interpret the presence of helium in the $\omega$ Cen giants and
diagram these ideas in Fig. 11.  
These authors evaluated three factors in presenting an evolution scenario: 
(1) morphological continuity on the CMD; (2) population fraction;
and (3) spectroscopic metallicity of the stars.   
The bifurcated main sequence (MS) principally represents two values 
of [Fe/H], $-$1.68 (spanning 0.2 dex) and $-$1.37,
and a minority population with [Fe/H] = $-$1.1 to $-$0.6.  Detailed abundance study of
the subgiant branch (SGB) reveals  (Villanova \etal\ 2007)  four or five components
which peak at specific values of [Fe/H] and the abundances typically span 
0.3 to 0.6 dex (FWHM) each, and range in [Fe/H] from $-$1.7 to
$-$0.6.  These too are marked in
Fig. 11 and denoted by SGB: A, B, C, and D.  Possibly an additional
component bridges the abundance range between B, C and D.  On the
red giant branch (RGB),  Johnson \& Pilachowski (2010) identified 4 components
which they denote (following Sollima \etal\ 2005), RGB: MP, MInt1,
MInt 2+3 and RGB-a where MP and MI stands for metal-poor and
metal-intermediate.  The abbreviation RGB-a denotes the anomalous 
red giant branch (Lee \etal\ 1999; Pancino \etal\ 2000). To understand the evolutionary relationships
among these components, Villanova \etal\ (2007) suggest 
'cross-over' occurs from the red main sequence (rms) to the
subgiant (SGB) components A, B, and C, and from the
blue main sequence (bms) to the subgiant subpopulations B and C,
and perhaps the intermediate component located between B, C, and D.

From the subgiant branch to the red giant branch, component SGB-A
feeds into the RGB-MP, and stars on SGB-D move to RGB-a. Thus the 
absence of helium in giants for [Fe/H]~$<$~$-$1.8 is not 
surprising following the route from the MS-rms to RGB-MP. The MS-rms
also overlaps
the metallicity of SGB-B,C and the component percentages suggest that
MS-rms evolves to SGB-B,C, and then to RGB-MInt1 (Villanova \etal\ 2007), 
The $\omega$~Cen stars with helium span [Fe/H] = $-$1.8 to $-$1.45 
in our sample. 
Villanova \etal\ (2007) conclude, based on photometric
and spectroscopic information that the MS-bms evolves through SGB-B,C
which
then, based on the metallicity overlap, contributes to  RGB-MP and
RGB-MInt1,
and perhaps RGB-MInt2+3.  In their abundance 
study of 855 red giants in $\omega$ Cen, 
Johnson and Pilachowski (2010) found that  O-poor, Na/Al rich
stars occur in three out of the 4 populations of red giants in the cluster,
but not in the RGB-a cohort.  The intermediate population of the [O/Fe] poor 
giants (RGB-MInt1 and RGB-MInt 2+3) shares attributes with the blue
main sequence: spatial distribution, metallicity, and percentages,
leading Johnson \& Pilachowski (2010) to suggest that the [O/Fe] 
depletion (coupled with the Na/Al enhancement) signals those 
giants whose progenitors belonged to the blue main sequence. The mix of
helium detections and non-detections between $-$1.8 and $-$1.45
appears consistent with the crossovers outlined above.

We are left with an apparently anomalous object, LEID 38269, which
is metal-rich, [Fe/H] = $-$1.16 and does not show helium. Its 
evolutionary path currently places it on the RGB-MInt2+3, and
its metallicity would seem to prevent an origin on the metal-poor MS-rms
where helium is weak or absent.   However, age dating of 
the subgiant population in $\omega$ Cen provides a clue to this object.

In addition to continuity of metallicity values, 
Villanova \etal\ (2007)  note that the subgiant branch offers an 
optimum region of the CMD to 
determine ages by comparison to stellar isochrones.   Interestingly, they find  the broadest
dispersion in stellar ages for stars with  $-1.8 \le [Fe/H] \le -1.5$
which echoes earlier results from studies of the main sequence
turnoff stars by Stanford \etal\ (2006). 
Over this metallicity range, 
the estimated ages  span almost a factor of two, from 0.6 of the 
age of the oldest stars  continuously ranging to the oldest stars in
the cluster.  Thus the mixed appearance
of helium in our targets at this metallicity range appears in 
harmony with the span of ages found in subgiants.  At the highest iron abundance
([Fe/H] $\sim$$-$1.1), three subgiants in the Villanova \etal\ (2007) sample have ages
commensurate with those of the most metal-poor stars suggesting a
simple age-metallicity correlation does not exist in $\omega$ Cen.  
This might occur as a result of 
several star-forming episodes.  Material
from Type II supernovae creates one stellar generation which is  
followed by the polluting effects from 
longer-lived metal-poor intermediate-mass asymptotic giant branch stars on
a subsequent stellar generation (Johnson \& Pilachowski 2010).
LEID 38269 also does not show enhanced Na and Al,  and 
would plausibly  belong to this older population. It
is the coolest and faintest star in our sample, and could form a part of 
the 'anomalous' population in $\omega$ Cen.

\subsection{Concluding Comments}

The direct detection of a helium abundance variation and its
correlations with both [Fe/H] and light element abundances
of Al and Na in this small sample of red giants,  supports  
the conjecture that helium is enhanced in the blue main sequence
component of the $\omega$~Cen cluster. It appears likely that the
helium-enhanced stars contain the products of high-temperature
hydrogen burning produced by  an earlier generation of stars.  
Spectra of additional targets
would be useful to thoroughly understand the relationship between
[Fe/H] and the presence of helium.

The infrared helium line at 10830\AA\ appears to be a powerful
probe of the presence of helium in warm giant stars. Detailed non-LTE
studies of its formation under a variety of conditions are needed to
be able to extend the interpretation to a larger range of temperatures
and luminosities, and also to extract the absolute abundance of
helium.

This research has made use of NASA's Astrophysics Data System. 
We thank the US Gemini Office for assistance in preparing Phase II, 
and providing  travel support to Gemini-S.   We thank Steve Margheim
and Claudia Winge  for Phase II advice and support during
our classical run at Gemini-S. JS acknowledges support provided 
by NASA through Hubble Fellowship grant HST-HF-51237.01, 
awarded by the Space Telescope Science Institute, which is 
operated by the Association of Universities for Research in Astronomy,
Inc., for NASA, under contract NAS 5-26555.

{\it Facilities:} \facility{Gemini:South
  (PHOENIX)}, \facility{Magellan/Clay (MIKE)}

\clearpage

\begin{figure}
\begin{center}
\includegraphics[angle=90, scale=0.7]{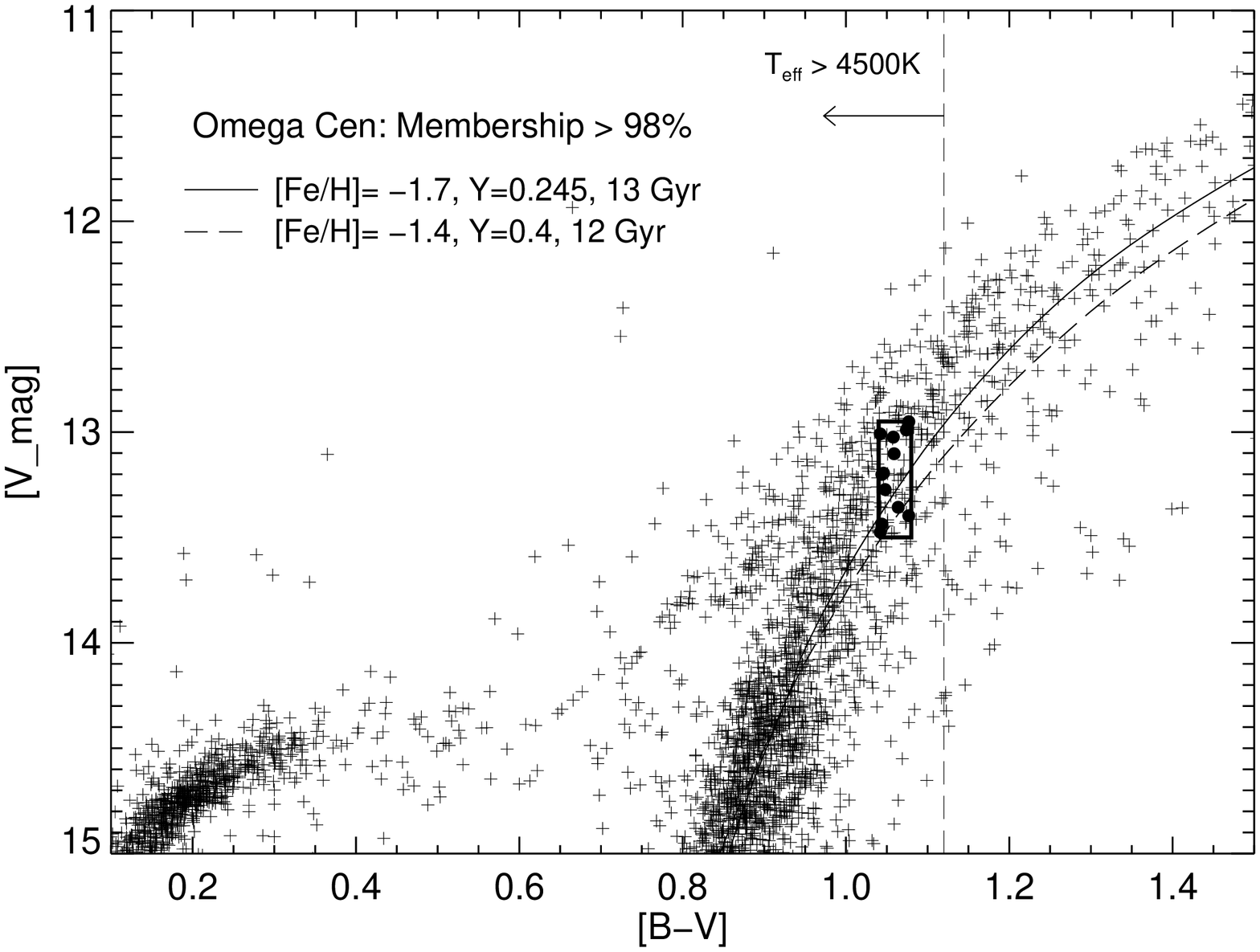}
\caption{Color-magnitude diagram  of $\omega$  Cen  stars from van Leeuwen et al. (2000) with
  better than 98\% probability of membership. Filled circles represent the
  stars observed. Note these targets are stars on the red giant
  branch; the target selection purposely avoids asymptotic branch objects.
  Isochrones are taken 
from the Dartmouth Stellar  Evolution
Database (http://stellar.dartmouth.edu/$\sim$models/) for the two majority
populations in Omega Cen (Norris 2004;Piotto \etal\ 2005) and taking E(B-V)=0.11 (Lub 2002).}
\end{center}
\end{figure}
\clearpage
\begin{figure}
\begin{center}
\includegraphics[angle=0.,scale=.9]{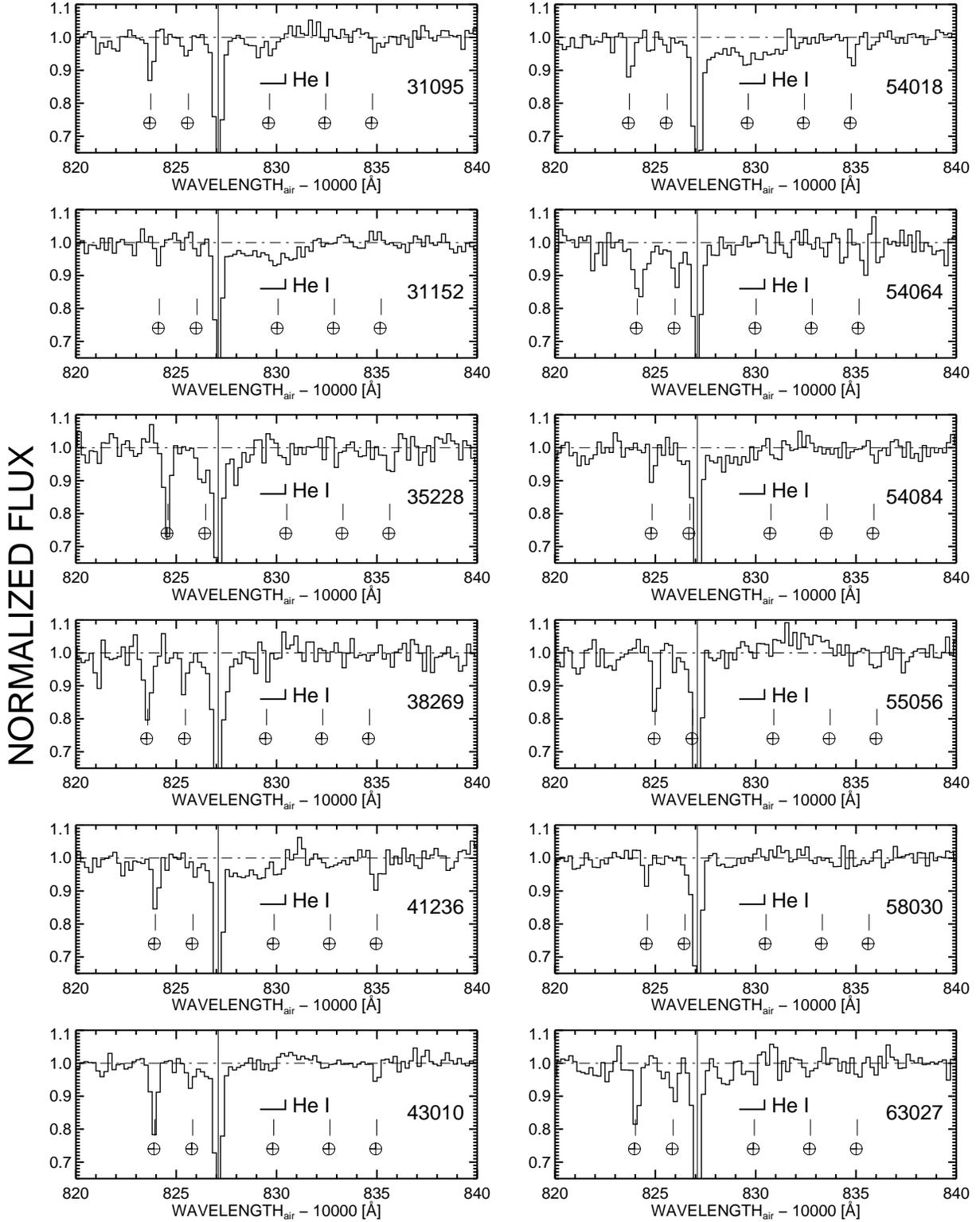}
\caption{Spectra from PHOENIX (binned to a resolution element) are
  aligned on the rest wavelength of the photospheric
  line of \ion{Si}{1} at $\lambda$ 10827.09. The LEID identification
  numbers are noted in each panel.  The position of the strongest
  helium
line component is marked by a vertical bar with an extent to shorter
wavelengths for the weak components.  Water vapor lines are noted by the $\oplus$ symbol.}

\end{center}
\end{figure}

\clearpage
\begin{figure}
\begin{center}
\includegraphics[angle=0., scale=0.95]{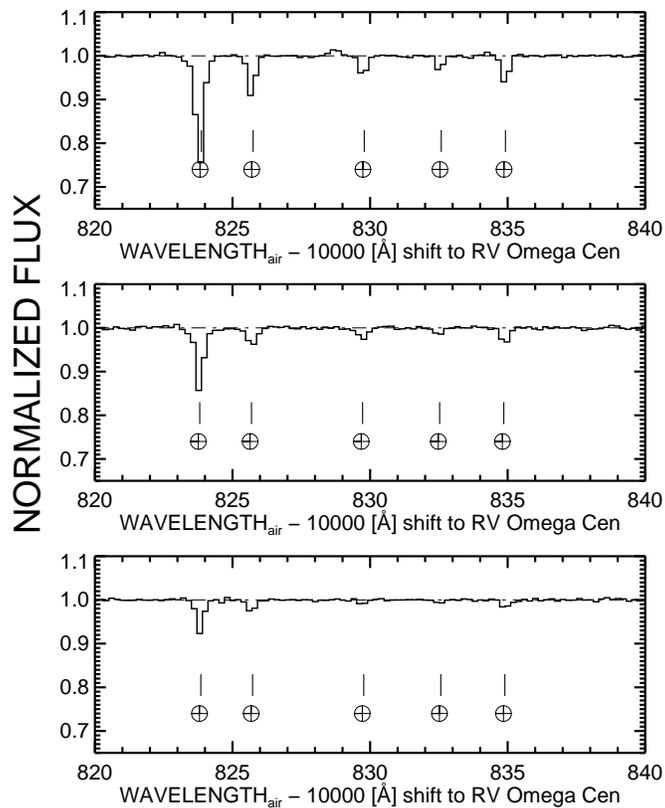}
\caption{Spectra of HR 5231, a fast rotating B2.5IV star showing
the variation of the strengths of the telluric water line.  These
spectra have been shifted by $-$230 \kms (8.3\AA) to correspond
to the average radial velocity of our target stars, and to make
comparison easier with Fig. 2.  The observed strength of the water
vapor
line near the helium feature (the central water vapor line near $\lambda$10830
in this figure) ranged from 4 to 18 m\AA\  in the target stars.  Wavelengths are taken from
Breckenridge \& Hall (1973).} 

\end{center}
\end{figure}
\clearpage
\begin{figure}
\begin{center}
\includegraphics[scale=0.7,angle=90]{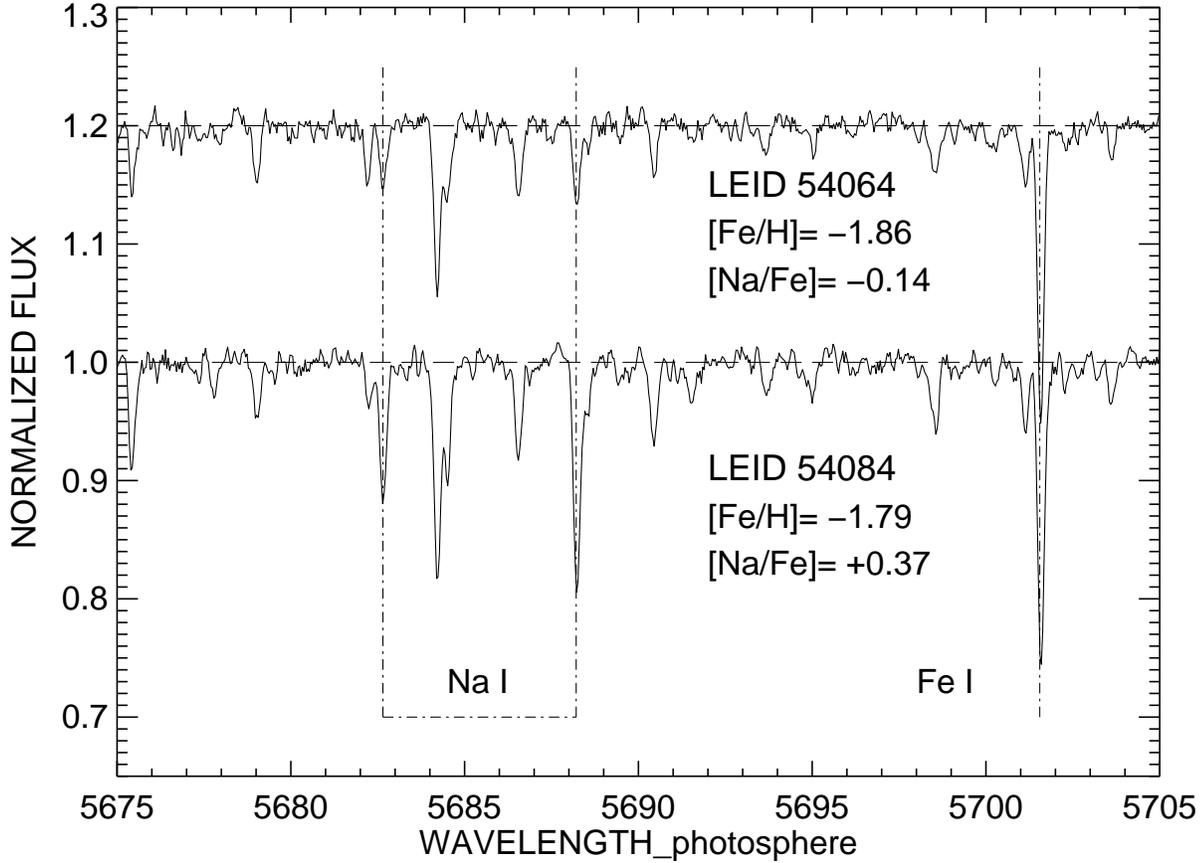}
\caption{Region of the Na lines (laboratory wavelengths)  in 
MIKE spectra of two stars: LEID 54064 and LEID 54084. They have 
similar values of [Fe/H], equivalent $T_{eff}$ and $log\ g$, and enhanced Na in LEID 54084, but not
in LEID 54064. LEID 54084 shows helium; LEID 54064 does not. The
spectrum of LEID 54064 has been offset by +0.2 in the normalized flux.}
\end{center}
\end{figure}
\clearpage

\clearpage
\begin{figure}
\begin{center}
\includegraphics[angle=90, scale=0.7]{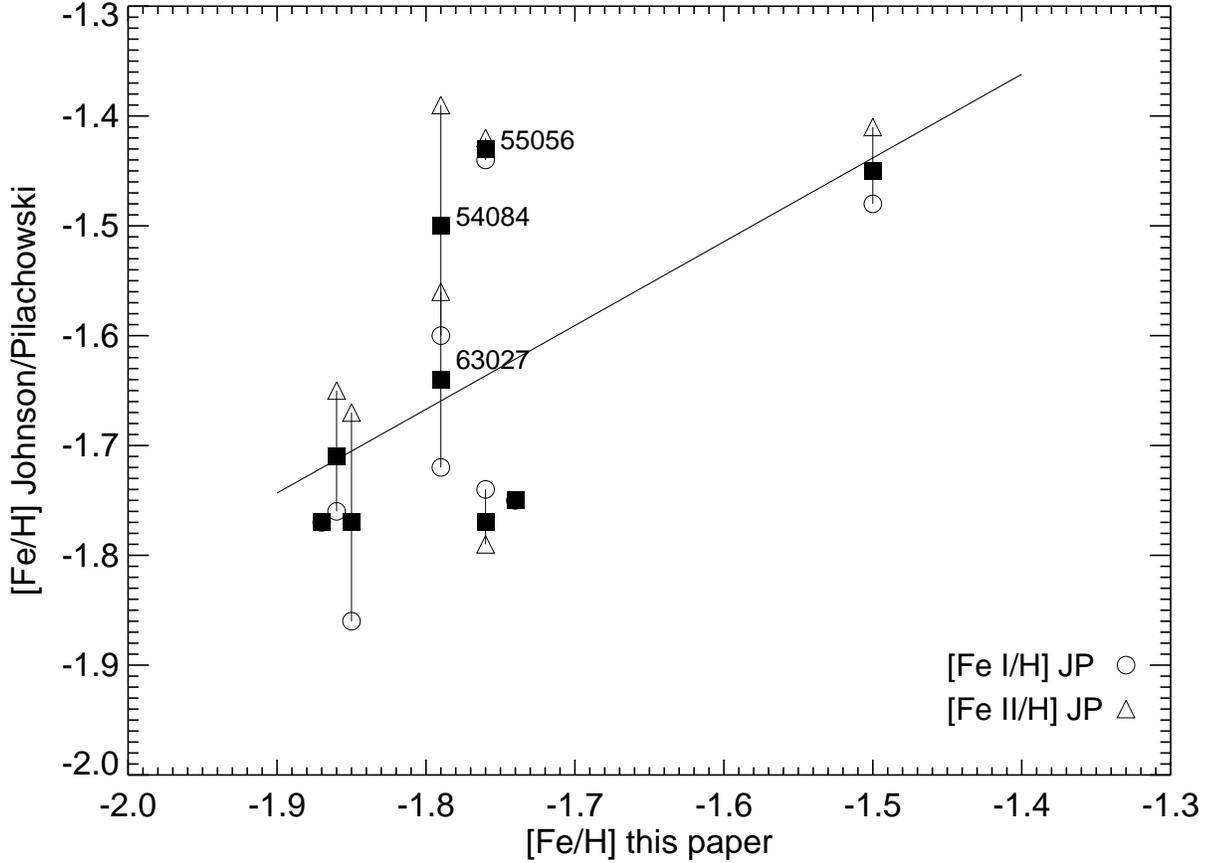}
\caption{Comparison of Johnson \& Pilachowski (2010) [Fe/H] abundances
from \ion{Fe}{1} ($\bigcirc${\it symbol}) and, in some cases one line or two of
\ion{Fe}{2} ($\triangle$ {\it symbol}), 
and the average value ($\blacksquare$~{\it symbol}) with the
abundances derived here from \ion{Fe}{2} lines in MIKE/Magellan echelle
spectra. The straight line marks a linear relation least-squares fit
where the mean difference between determinations amounts to  $\sim$$-$0.12$\pm$0.12 dex.
}
\end{center}
\end{figure}

\clearpage

\begin{figure}
\begin{center}
\includegraphics[angle=90.,scale=0.7]{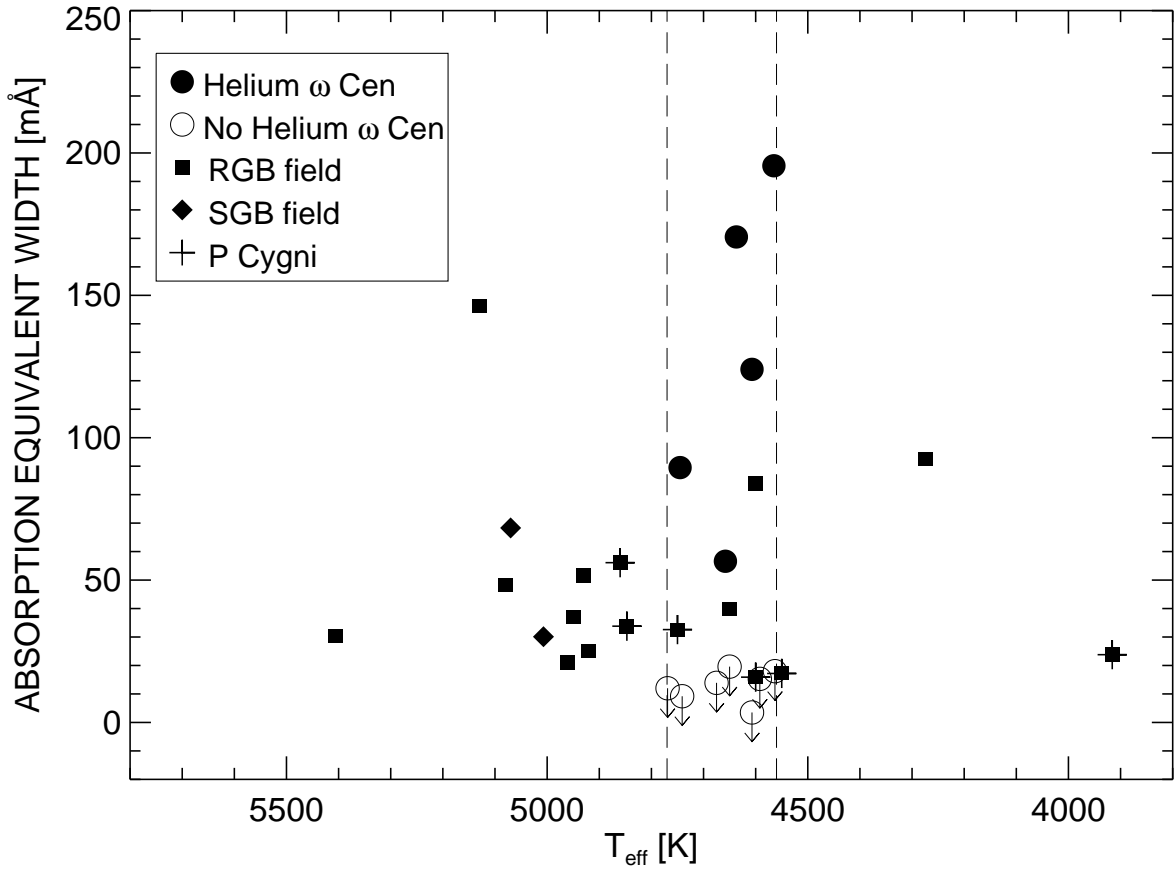}
\caption{Helium equivalent width measures for the $\omega$ Cen giants
  ({\it this paper}), and  metal-poor red giants and 
subgiants in the field (from Dupree \etal\ 2009).  The broken 
vertical lines mark the narrow temperature region of the
$\omega$ Cen sample.}

\end{center}
\end{figure}

\clearpage


\begin{figure}
\begin{center}
\plottwo{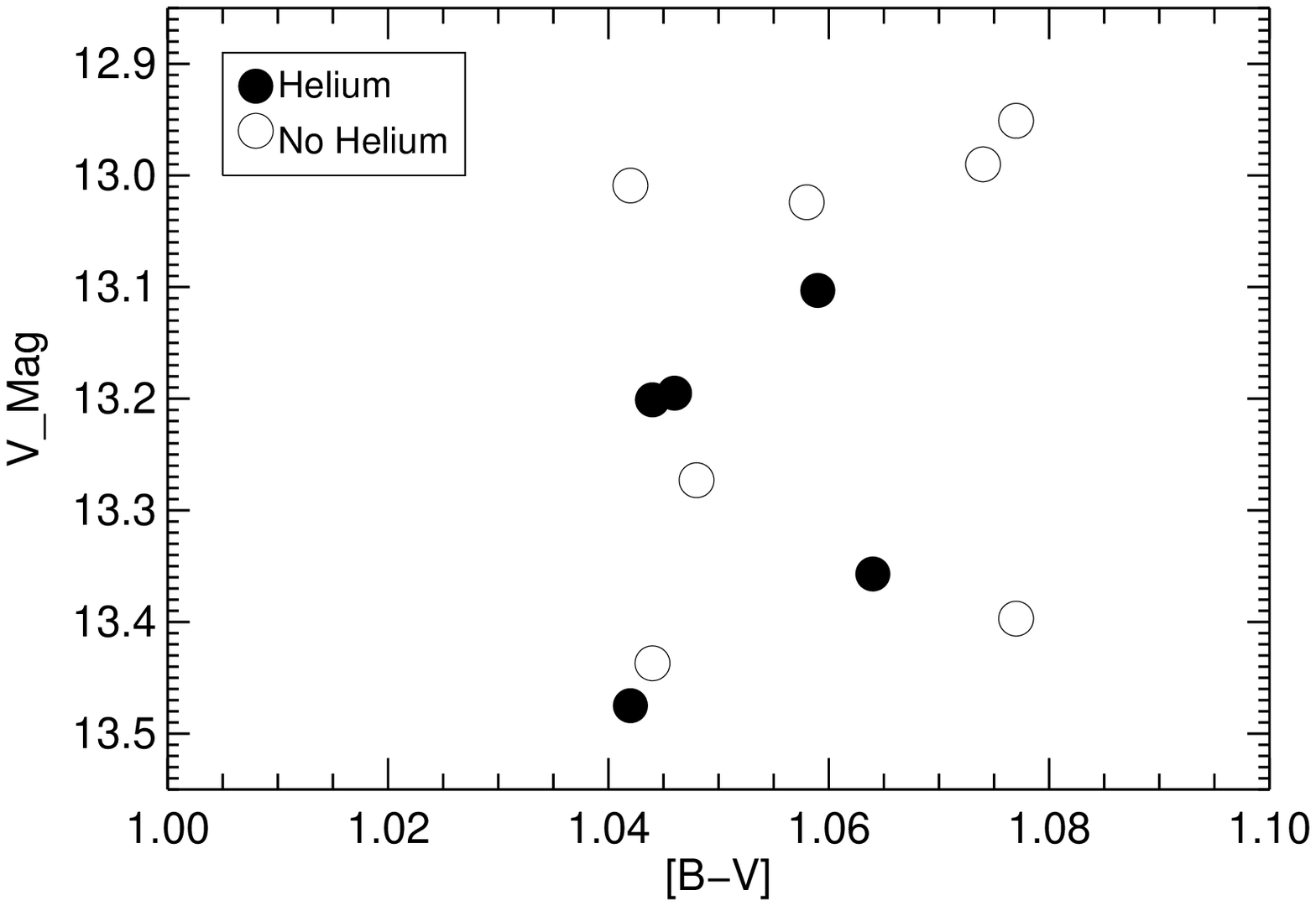}{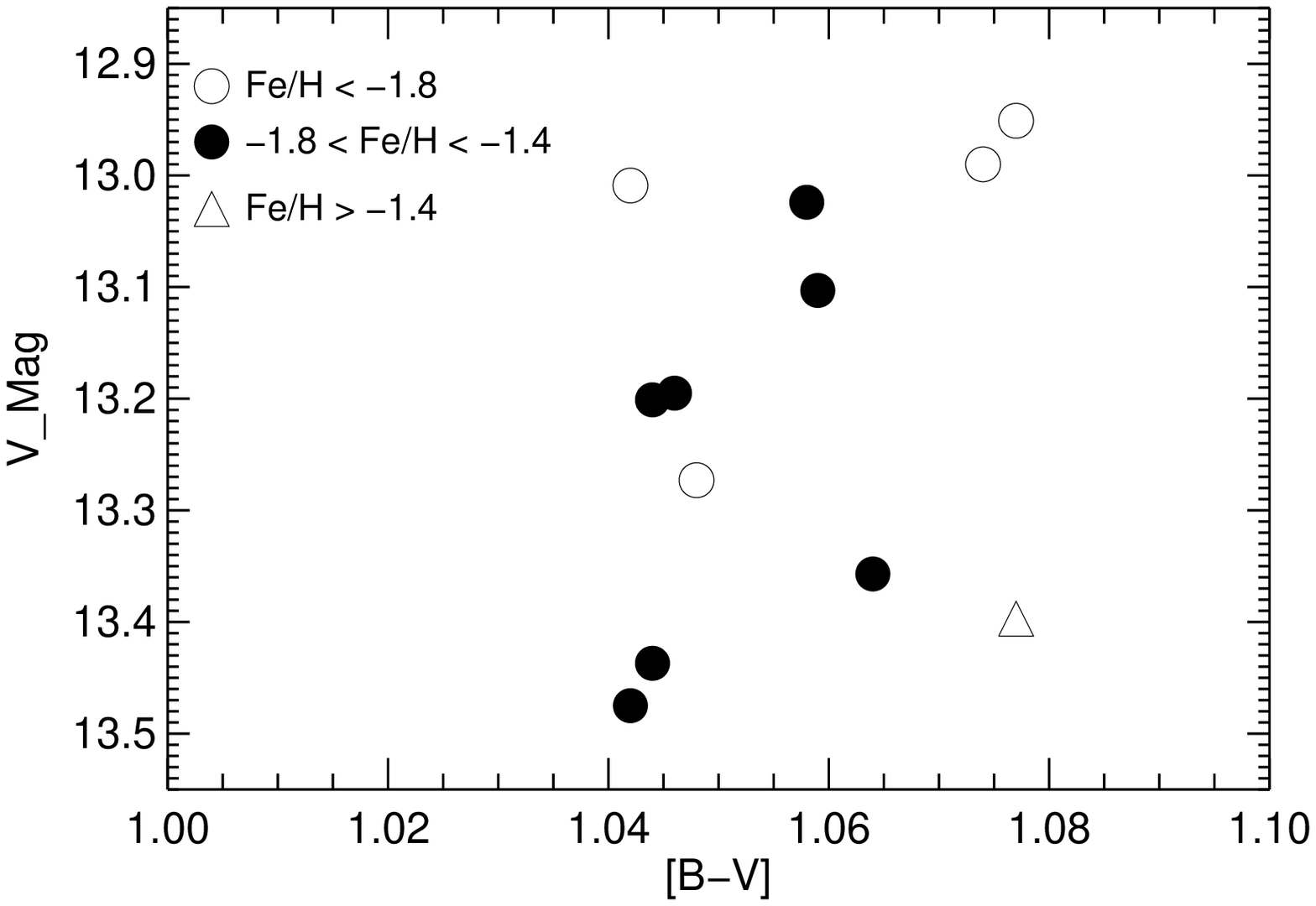}
\caption{The color-magnitude diagram for the $\omega$ Cen stars
  observed
with PHOENIX. In the {\it left panel}, the stars are labeled
  according to the presence of a \ion{He}{1} line in the spectra. In
the {\it right panel}, stars are labeled according to  [Fe/H] abundance.}

\end{center}
\end{figure}

\clearpage
\begin{figure}
\begin{center}
\includegraphics[angle=0,scale=0.9]{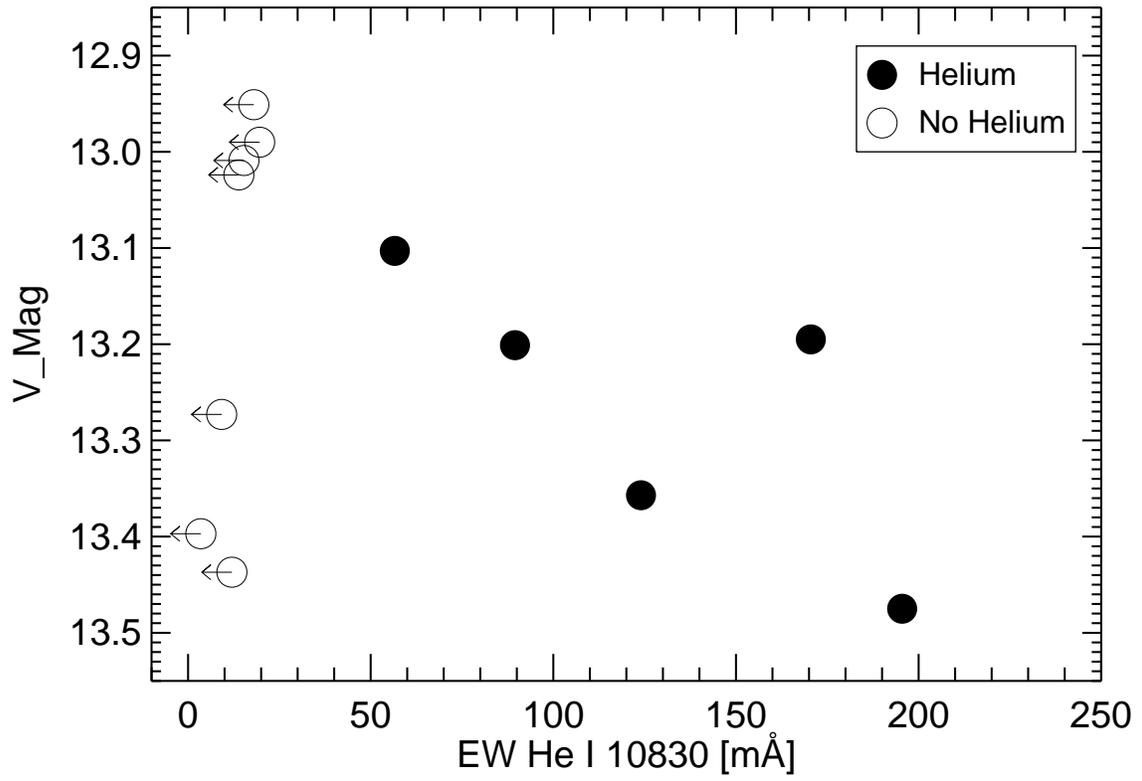}
\caption{Relation between the helium equivalent width and apparent
  magnitude of our targets suggests the line may become stronger in 
the less luminous stars. The  non-detections can be considered upper
  limits
to the helium equivalent widths.}
\end{center}
\end{figure}


\clearpage
\begin{figure}
\begin{center}
\includegraphics[angle=0,scale=0.9]{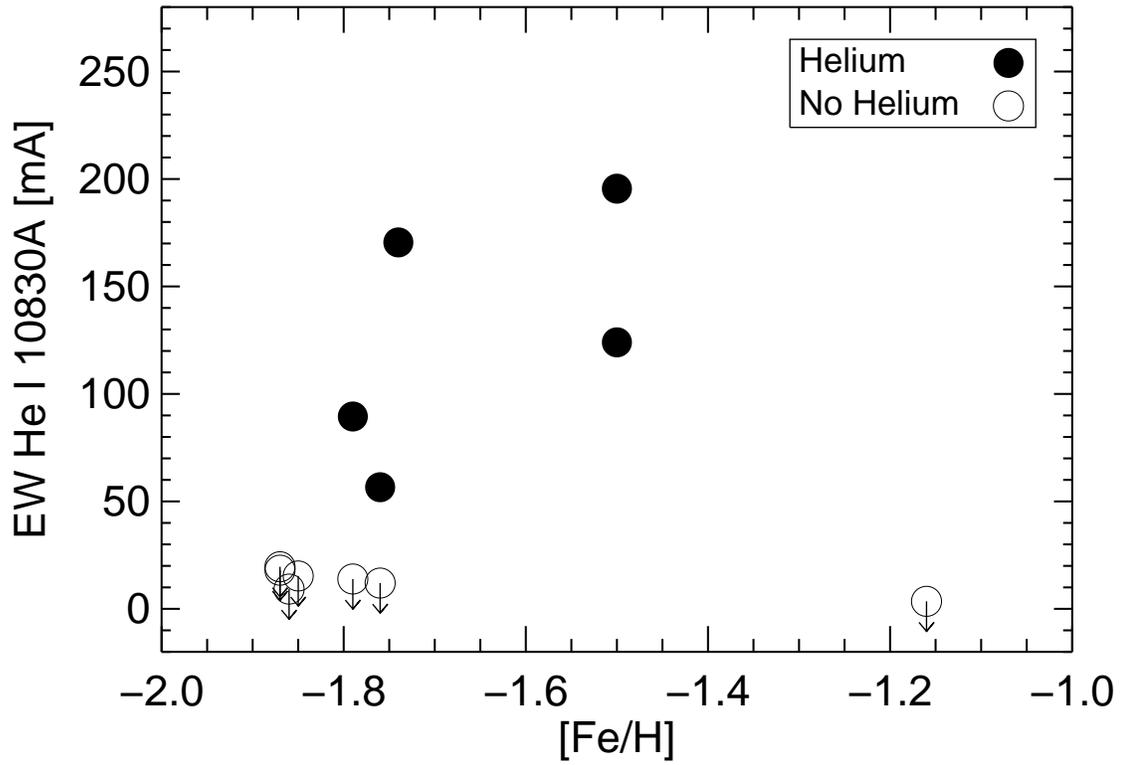}
\caption{Dependence of equivalent width of the helium line on the
  value of [Fe/H]. The equivalent widths for the non-detections,
  marked 
as upper limits, represent the widths of the water vapor lines near the wavelength of
helium in the target stars.}

\end{center}
\end{figure}
\clearpage

\begin{figure}
\begin{center}
\includegraphics[angle=0,scale=0.9]{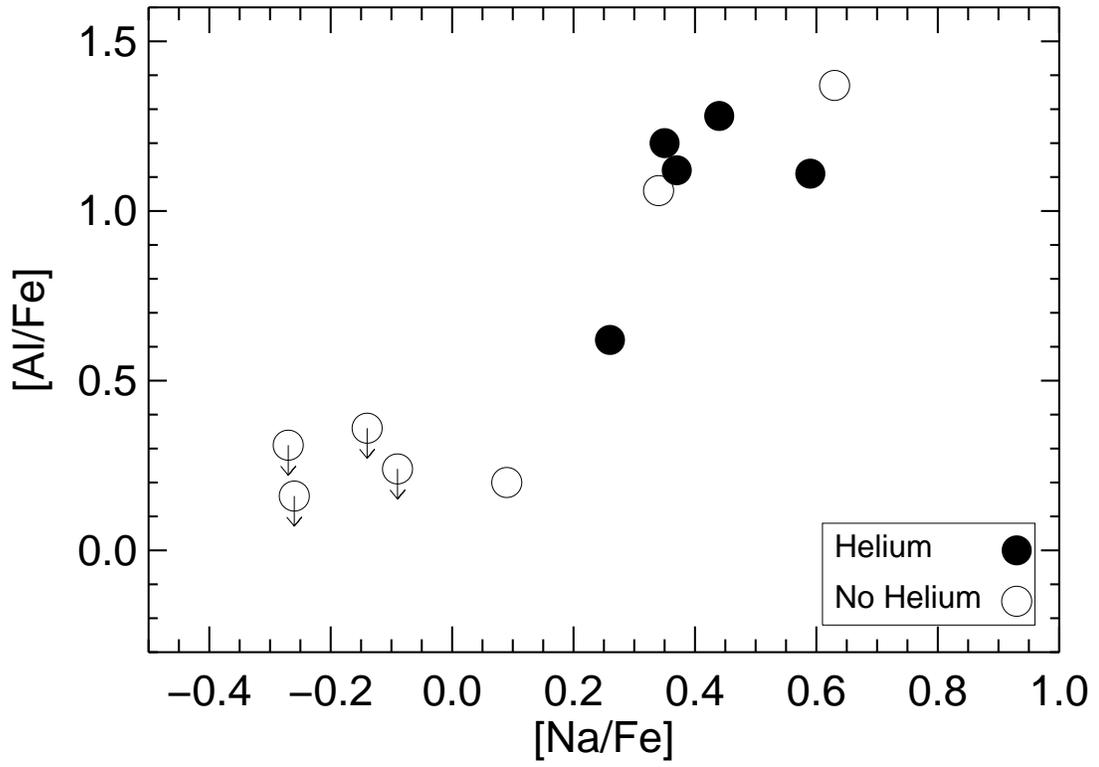}
\caption{Relation between [Al/Fe] and [Na/Fe] where the stars with
  detected helium are preponderantly those with enhanced Al and Na.}

\end{center}
\end{figure}

\clearpage


\begin{figure}
\begin{center}
\hspace*{-0.5in}
\includegraphics[angle=0,scale=0.75]{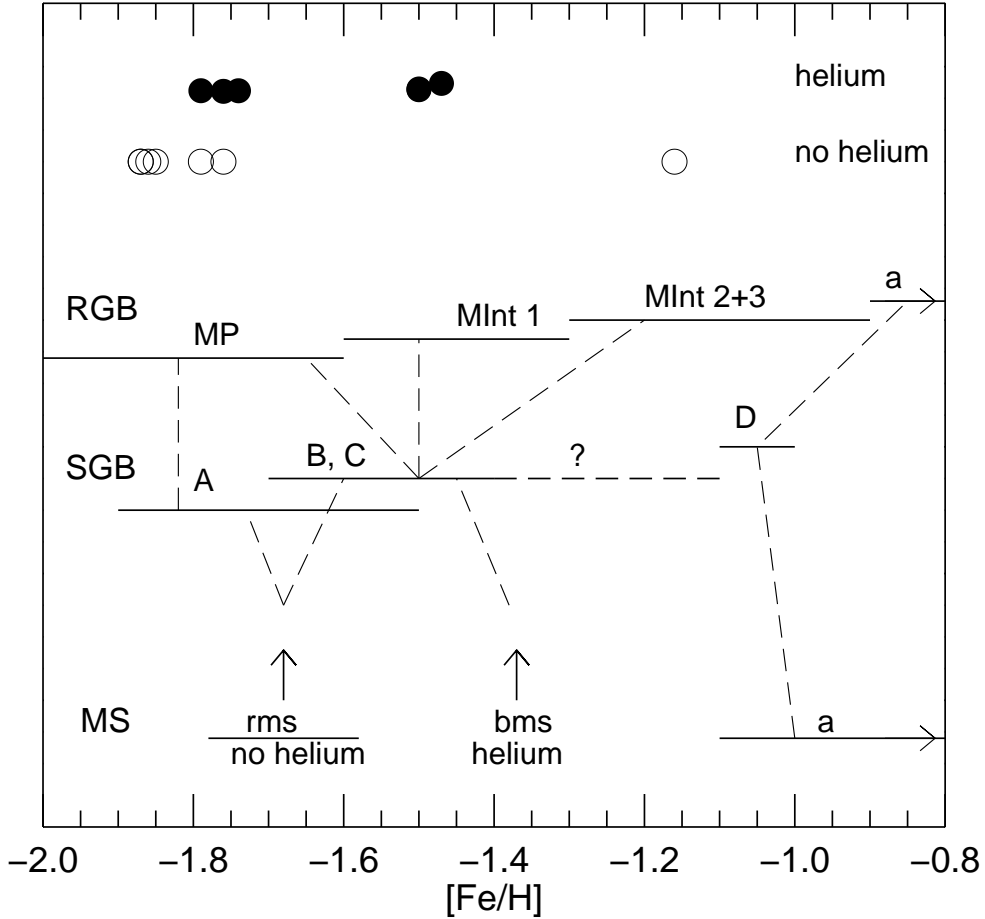}
\caption{A posited evolutionary
  sequence of the populations in $\omega$ Cen in relation to the 
presence or not of helium in the red giants reported here (indicated
  by open or filled circles).  The [Fe/H] abundances 
of population components for the main sequence (Piotto \etal\ 2005),
the sub-giant branch, SGB (Villanova \etal\ 2007) and the
red giant branch RGB, (Johnson \& Pilachowski 2010) are marked. 'Cross-over'
among the components, has been suggested by Villanova et al (2007) as
possibly occurring from the red main sequence (rms) to the
subgiant (SGB) components A, B, and C as well as from the blue main sequence
(bms) to the subgiant subpopulations B and C. Johnson \& Pilachowski's results
  (2010) also support the bms as the progenitor of the MInt1 and
  MInt2+3 RGB population.  Ages are not considered
  here; the estimate by Villanova \etal\ (2007) suggests that the stars
  without a helium detection belong to the oldest stellar generation in
  $\omega$~Cen (see text). }
\end{center}
\end{figure}

\clearpage


\begin{deluxetable}{rrrrrcrcrrr}
\def\a{\phantom{0}}
\def\b{\phantom{00}}
\tablecolumns{11}
\rotate
\tablewidth{0pt}\tablenum{1}
\tablecaption{Omega Cen Giants Observed}
\tablehead{
\colhead{Star\tablenotemark{a}}   &
\colhead{$V$\tablenotemark{a}}  &
\colhead{$B-V$\tablenotemark{a}}  &
\colhead{RV} &
\colhead{EW } &
\colhead{He I }&
\colhead{T$_{eff}$} &
\colhead{log g} &
\colhead{[Fe/H]}  &
\colhead{[Na/Fe]} &
\colhead{[Al/Fe]}\\
\colhead{}&\colhead{}&\colhead{}&\colhead{[km s$^{-1}$]}&\colhead{[m\AA]}&\colhead{}&
\colhead{[K]}&\colhead{[cm s$^{-2}$]}&\colhead{}&\colhead{}&\colhead{}}
\startdata
LEID 31095&13.10&1.059  &$+$244.5&56.6&yes&4658&1.65&$-$1.76&0.35&1.20 \\
LEID 31152&13.20&1.046  &$+$234.9&171&yes&4637&1.67&$-$1.74&0.26&0.62     \\
LEID 35228&12.95&1.077  &$+$224.7&$\le$18 &no&4563&1.54&$-$1.87&0.34&1.06     \\
LEID 38269&13.40&1.077  &$+$250.5&$\le$3.5 &no&4607&1.74&$-$1.16&0.09&0.20  \\
LEID 41236&13.36&1.064  &$+$238.4&124&yes&4607&1.72&$-$1.50&0.44&1.28   \\
LEID 43010&13.01&1.042  &$+$241.8&$\le$15.3 &no& 4592&1.57&$-$1.85&$-$0.26&$\le $0.16 \\
LEID 54018&13.48&1.042  &$+$246.1&196&yes& 4565&1.74&$-$1.50&0.59&1.11 \\
LEID 54064&13.27&1.048  &$+$233.6&$\le$9.2 &no& 4741&1.76&$-$1.86&$-$0.14&$\le$0.36 \\
LEID 54084&13.21&1.044  &$+$213.8&89.5&yes& 4745&1.74&$-$1.79&0.37&1.12 \\
LEID 55056&13.44&1.044  &$+$208.6&$\le$12 & no&4769&1.84&$-$1.76&0.63&1.37\\
LEID 58030&12.99&1.074  &$+$223.8&$\le$19.6&no&4650& 1.60&$-$1.87&$-$0.27&$\le$0.31\\
LEID 63027&13.02&1.058  &$+$237.1&$\le$13.9&no&4675&1.63&$-$1.79&$-$0.09&$\le$0.24 \\

\enddata
\tablenotetext{a}{Identification numbers, magnitudes and colors taken from van Leeuwen et al. (2000).}

\end{deluxetable}


\begin{deluxetable}{llccccccccccccc}
\def\a{\phantom{0}}
\def\b{\phantom{00}}
\tablecolumns{15}
\rotate
\tabletypesize{\scriptsize}
\tablewidth{0pt}\tablenum{2}
\tablecaption{Atomic Parameters and Equivalent Widths\tablenotemark{a}}
\tablehead{
\colhead{Ion}   &
\colhead{$\lambda$ [\AA]}  &
\colhead{log $gf$}  &
\colhead{31095} &
\colhead{31152} &
\colhead{35228}&
\colhead{38269} &
\colhead{41236} &
\colhead{43010}  &
\colhead{54018} &
\colhead{54064}&
\colhead{54084}&
\colhead{55056}&
\colhead{58030}&
\colhead{63027\tablenotemark{b}}}
\startdata
Fe II&4620.52&$-$3.23&38.3&30.1&36.3&51.8&42.2&35.5&41.1&32.7&35.7&39.2&31.9&36.4 \\
Fe II&4670.17&$-$4.05&14.0&15.3& 16.7&23.9&15.9&13.5&19.7&12.9&14.8&10.3&14.3&18.7\\
Fe II&5100.66&$-$4.20&9.1&\nodata&\nodata&25.0&18.0&7.5&12.9&9.7&8.5&8.9&\nodata&9.5 \\
Fe II&5132.67&$-$4.01&9.6 &8.2&8.6&25.0&17.1&9.9&11.1&8.2&9.9&9.7&\nodata&11.0\\
Fe II&5197.58&$-$2.24&71.7&65.0&72.9&83.1&74.3&72.3&72.9&65.6&70.6&74.3&66.3&69.8\\
Fe II&5234.63&$-$2.10&75.1&70.7&73.7&81.7&79.2&71.9&74.3&71.4&70.0&75.3&70.7&73.9\\
Fe II&5264.79&$-$3.02&28.1&22.6&25.3&39.0&28.9&24.2&32.4&23.7&24.9&29.3&24.5&21.6\\
Fe II&5991.38&$-$3.57&15.0&17.1&15.3&30.4&18.1&16.2&19.3&15.6&16.7&16.6&16.3&15.7\\
Fe II&6084.10&$-$3.81&10.9&11.7&3.9&18.9&14.1&7.3&15.7&7.5&9.5&13.6&8.4&9.1\\
Fe II&6416.93&$-$2.68&15.3&13.9&15.0&22.7&20.2&14.7&20.6&13.0&15.0&13.2&10.3&14.3\\
Fe II&6432.68&$-$3.57&26.9&24.0&23.8&35.7&25.3&24.5&24.0&22.6&27.0&25.4&24.0&25.5\\
Fe II&6456.39&$-$2.13&48.4&38.6&42.5&49.2&50.6&38.2&46.4&37.0&40.4&46.7&39.3&42.4\\
Fe II&6516.08&$-$3.28&40.6&34.3&38.3&51.7&41.3&36.7&39.8&34.3&39.3&38.6&33.7&36.0\\
\hline
Al I&6696.03&$-$1.45&41.1&12.1&29.9&16.1&61.7&5.9&46.7&6.8&30.8&43.8&7.7&4.9\\
Al I&6698.67&$-$1.87 &20.4&7.5&12.5&9.8&35.9&\nodata&30.3&2.1&14.5&23.4&2.9&4.6\\
Al I&7835.32&$-$0.74&30.2&11.7&19.4&19.7&48.5&2&39.6&4.2&19.6&37.1&1.9&6\\
Al I&7836.13&$-$0.45&31.0&13.3&26.8&18.7&57.3&\nodata&53.3&3.7&28.8&44.7&\nodata&1.8\\
\hline
Na I&4668.57&$-$1.41&8.0& 7.0&7.5&6.1&19.6&3.1&18.1& 3.7&4.6&8.6&\nodata&\nodata\\
Na I&5682.65&$-$0.70&36.9&30.4&27.2&54.9&60.3&13.2&78.7&12.5&29.5&44.9&10.3&15.6  \\
Na I&5688.21&$-$0.37&57.5&43.2&50.9&74.1&75.5&17.0&93.3&15.8&45.7&68.4&16.3&22.8  \\
Na I&6154.23&$-$1.56&8.7&$<$10.7\tablenotemark{c}&6.8&17.9&14.4&$<$5.1&19.0&2.2&4.9&$<$13.6&3.5&$<$5 \\
Na I&6160.75&$-$1.26&12.5&13.6&13.5&26.6&25.7&5.0&30.2&5.3&12.6&22.6&2&$<$4\\
\enddata
\tablenotetext{a}{Stars denoted by LEID number (van Leeuwen et
  al. 2000); Wavelengths and $gf$ values from Fulbright (2000) except
  for
Al I which were taken from Jacobson \etal\ (2009);
Equivalent widths given in m\AA.}
\tablenotetext{b}{Na abundance for LEID 63027 was also derived from
  the
resonance lines ($\lambda$ 5889.96 and $\lambda$5895.94) and the
abundance is in harmony with the values derived from the two Na lines
listed below.}

\tablenotetext{c}{The $<$ symbol indicates the feature is blended,
  difficult to separate, and
  the value represents an upper limit to the equivalent width of the line.}

\end{deluxetable}


\end{document}